\documentclass[12pt, letterpaper]{JHEP3}

\usepackage{amssymb, amsmath, amsopn, amsthm}
\usepackage{epsfig}

\newcommand{\spacer}{\phantom{spacer}}
\newcommand{\contract}{\llcorner}
\newcommand{\barray}{\begin{eqnarray}}
\newcommand{\earray}{\end{eqnarray}}
\newcommand{\re}{{\rm Re}\,}
\newcommand{\im}{{\rm Im}\,}

\newcommand{\cD}{\mathcal{D}}

\newcommand{\cF}{\mathcal{F}}

\newcommand{\bZ}{\mathbb{Z}}

\newcommand{\td}{\tilde}
\renewcommand{\Re}{{\rm Re}\,}
\renewcommand{\Im}{{\rm Im}\,}

\newcommand{\beq}{\begin{equation}}
\newcommand{\eeq}{\end{equation}}
\newcommand{\tc}{\td{c}}
\newcommand{\er}[1]{(\ref{eq:#1})}
\newcommand{\F}{\cF}
\newcommand{\infinity}{\infty}

\title{Generalized Flux Vacua}

  \author{Jessie Shelton$^1$, Washington Taylor$^{1,2}$, and Brian Wecht$^1$ \\
{${}^1$Center for Theoretical Physics} \\ {Massachusetts Institute of Technology} \\
{Cambridge, MA 02139, U.S.A.} \\
{$\,$} \\
{${}^2$Department of Physics}\\
{Stanford University}\\
{Stanford, CA 94305--4060, USA}\\
{$\,$} \\
{\tt jshelton\ {\rm at}\ mit.edu, wati\
  {\rm at}\ mit.edu, bwecht\ {\rm at}\ mit.edu}}

\abstract{We consider type II string theory compactified on a
symmetric $T^6/\bZ_2$ orientifold.  We study a general class of
discrete deformations of the resulting four-dimensional supergravity
theory, including gaugings arising from geometric and ``nongeometric'' fluxes, as well as the usual R-R and NS-NS fluxes.  Solving the equations of
motion associated with the resulting ${\cal N} = 1$ superpotential, we
find parametrically controllable infinite families of supersymmetric
vacua with all moduli stabilized.  We also describe some aspects of
the distribution of generic solutions to the SUSY equations of motion
for this model, and note in particular the existence of an apparently
infinite number of solutions in a finite range of the parameter space of the four-dimensional effective theory. }

\preprint{hep-th/0607015, MIT-CTP-3756, SU-ITP-06/16}

\begin{document}
\tableofcontents

\section{Introduction}
\label{sec:introduction}

Classifying the full space of string theory vacua is an enormous
problem.  For any given string theory, there is a broad range of
possible compactification geometries.  Imposing supersymmetry
simplifies the problem somewhat, and for many years attention was
focused on Calabi-Yau manifolds as the primary structure needed for
supersymmetric compactification of string theory.  In recent years,
however, more and more attention has focused on compactifications
including fluxes of various $p$-form fields on topologically
nontrivial cycles of the compactification manifold
\cite{flux-c,gvw,drsgkp}.  This has led to the realization that a broader
class of manifolds are needed even for supersymmetric
compactifications \cite{glmw,Hitchin}.  It is possible, however, that concrete realizations of compactifications explored to date only scratch the surface of the full range of possibilities.  In general, supersymmetric compactification on a
Calabi-Yau or other related manifold with fluxes leads to a gauged
supergravity theory in four dimensions.  In many gauged
supergravity theories, the parameters associated with the gauging of
the supergravity can be identified in this way with fluxes on a compactification
manifold or with certain features of the geometry of the
compactification manifold \cite{ss,km,x1}.  In other cases, however,
it is not known how to interpret the gauging parameters in ten
dimensions. It is possible that in some cases such gaugings are the result of
compactifications on nongeometric string backgrounds \cite{stw,
Dabholkar:2005ve}. Thus, while
many interesting four-dimensional solutions, including stable de Sitter vacua
\cite{vp}, may be realized in effective four-dimensional supergravity
theories, we often do not know how to interpret these solutions in
terms of ten-dimensional string theory.

When two ten-dimensional theories are related through a duality
symmetry, such as T-duality or mirror symmetry, we have a situation
where a given four-dimensional effective theory may be derived in two
different ways.  In this case, gauging parameters which have a natural
interpretation as fluxes in one picture may not have a known
interpretation in the dual picture.  The existence of one picture in
which these gauging parameters can be understood in ten dimensions,
however, suggests that these parameters should be allowed to take
nonzero values in a more general context, and that it may be possible
to turn on arbitrary combinations of these parameters even when there
is no single picture in which they all have a known interpretation.
This observation has motivated analysis of the duality transformation
properties of the gauging parameters in four-dimensional supergravity
theories \cite{ferrara}.  In \cite{stw}, we considered a simple
example of this situation, namely, a toroidal orientifold in Type II
string theory.  We derived the four-dimensional superpotential for
this toroidal orientifold, including NS-NS and R-R fluxes in both the
IIA and IIB theories.  These fluxes give rise to different subsets of
the set of possible gaugings of the four-dimensional model.  We argued
that to have a class of models which are invariant under duality, it
was natural to include a full duality-invariant set of gauging
parameters.  When we consider T-duality transformations on the
toroidal orientifold, the duality-invariant set of gauging parameters
includes integers which appear as T-duals of NS-NS 3-form flux $H$.
These integers can be
interpreted as parameterizing twists on the torus, often called
``geometric flux'' (as can be found after one T-duality using the Buscher rules \cite{Buscher}) and as generalized fluxes
(after two or three T-dualities).
\footnote{Here the word ``generalized" is not used in the sense of
``generalized geometry" \cite{Hitchin}.  We use it instead to describe
essentially topological structures which are related to $p$-form flux
under duality.}

These generalized fluxes can be referred to as ``nongeometric,'' in
the sense that when they are present in the absence of other fluxes,
they describe backgrounds with no global (or possibly even local)
geometric interpretation.  There is so far no complete worldsheet
description of spaces with these fluxes as string backgrounds, but one
proposal for a stringy construction of these fluxes has been described
by Hull \cite{x2} and explored by several authors at both the
classical \cite{hre, global,Lawrence:2006ma} and quantum
\cite{quantum} levels. While simple combinations of geometric and
generalized fluxes may have a string interpretation along these lines,
when sufficiently complicated combinations of fluxes are included, we
have as yet no concrete idea of how to lift the resulting
four-dimensional physics to a full string compactification.  While we
touch on this issue here, for the most part we simply treat the fluxes
as parameters in the four-dimensional effective theory.  Other
approaches to the worldsheet description of nongeometric backgrounds
include asymmetric orbifolds \cite{ao,ao2}, and a recent
generalization \cite{Hellerman:2006tx}.  We do not expect, however,
that the backgrounds described in the present work have an asymmetric
orbifold description.

In this paper we consider the four-dimensional theory with this
general set of fluxes and solve the tree-level supergravity equations.
We find classes of vacua in which the string coupling $g$ and
cosmological constant $\Lambda$ are parametrically controlled.  These
vacua are reminiscent of the parametrically controlled IIA vacua
described in \cite{dgkt2}, although in that case there was an explicit
picture of the vacua as ten-dimensional compactifications on a large
radius Calabi-Yau manifold.  Here, we do not have a ten-dimensional
picture of how these vacua should be interpreted, and likely have no
control over $\alpha'$ corrections.  Indeed, the structure of the
fluxes suggests that to construct a complete lift of these solutions
to a string compactification may require the introduction of
additional light fields.  Because the fluxes capture some essentially
topological features of the theory, however, we believe that the
existence of these parametrically controlled SUSY vacua points to some
new class of either geometric or nongeometric string vacua which we
may understand more clearly in due course.

In addition to demonstrating parametrically controlled vacua, we will
also study the broad features of the general supersymmetric solutions
in this model.  The generic solutions stabilize all moduli at tree
level, and are not equivalent to known purely geometric compactifications in
any duality frame.  We find that our toroidal model possesses an
apparently infinite number of (tree-level) solutions in finite ranges
of parameter space. This somewhat surprising result seems to 
contradict recent predictions regarding
properties of the string landscape
\cite{Acharya:2006zw,Ooguri:2006in}, though as we will discuss there
are some reasons why the solutions we find may not correspond to
stable nonperturbative vacua in a complete string theory framework.

In Section 2 we review the construction of the superpotential for the
dimensionally reduced four-dimensional theory on the toroidal
orientifold, including generalized fluxes.  We give a suggestive new
concise description of the constraints on consistent generalized
fluxes in terms of a generalized derivative operator.  In Section 3 we
give the equations of motion and briefly discuss some simple classes
of solutions.  In Section 4 we describe two two-parameter families of
parametrically controlled SUSY vacua, and in Section 5 we briefly
describe the statistics of more general vacua.  Section 6 contains
some concluding remarks.

\section{Review and Preview}
\label{sec:review}
\subsection{Generalized Fluxes}

In \cite{stw} we identified a set of NS-NS ``fluxes'' on the torus, which can individually be
obtained from the usual three-form $H$-flux by T-duality. T-duality
acts on the integrated NS-NS fluxes according to the chain
\beq \bar H_{abc} \stackrel{T_a}{\longleftrightarrow} f^a_{bc}
\stackrel{T_b}{\longleftrightarrow} Q^{ab}_c
\stackrel{T_c}{\longleftrightarrow} R^{abc}.
\label{eq:t-chain}
\eeq
Here $\bar H_{abc}$ is the integer number of units of $H$-flux on the
$(abc)$-cycle of the torus.  The ``geometric flux'' $f ^ a_{bc}$
characterizes the field strengths of a basis of one-forms $\eta^a$ satisfying
\beq
\label{eq:fdef}
 d\eta^a = f ^ a_{bc}\, \eta ^ b\wedge\eta ^ c.  
\eeq 
Differentiating \er{fdef} yields a Jacobi identity, and thus the $f ^
a_{bc}$ may be viewed as the structure constants for a Lie algebra.
The one-forms $\eta^a$ can be related to both the generators of the
isometry group of the compactification manifold and to a basis for its
frame bundle. When the Lie algebra defined by the $f ^ a_{bc}$ is
nilpotent, the $\eta^a$ can be constructed explicitly, and the
geometric flux can be straightforwardly interpreted as adding twists
to an underlying torus.  When the algebra is not nilpotent, then the
compact space is less closely related to the torus, and from the
four-dimensional point of view the lift to ten dimensions is more
subtle \cite{km}; we will return to this point below.

Following the T-duality chain of equation \er{t-chain}, acting on
$f^a_{bc}$ with
an 
additional
T-duality in the direction $b$ takes us to the nongeometric structure
characterized by $Q ^{ab}_c$.
\footnote{Strictly speaking,
the isometry in the direction $b$ is not globally
defined, but it seems that 
the T-duality in this direction can be understood, at least in the
dimensionally reduced theory, in terms of an orbifold by a combination
of a shift and a T-duality symmetry; see also \cite{global}.}
For a torus with $N$ units of such a $Q$-flux, and no other NS-NS
fluxes, fields are periodic only up to an identification which mixes
geometric data with the (integrated) $B$-field, and R-R $p $-forms of differing
degree with each other.  While the resulting compactification has a
geometric description locally, the transition functions render the
global description nongeometric.

Finally, the last step in the T-duality chain is a formal T-duality on
$c$, taking us to an object $R ^{abc}$.  This is the NS-NS analogue of
the R-R T-duality rule which takes $\bar F_x \stackrel{T_x
}{\longleftrightarrow} F_0$. A compactification with $R$-flux has
apparently no geometric description, even locally. A heuristic
argument for this lack of local geometry goes as follows: Consider a
$T^3$ with nonzero $H$ flux.  It is not possible to wrap a D3-brane on
this $T^3$ for the simple reason that turning on $H$-flux would give a
nonzero Bianchi identity for the gauge field living on the D3-brane,
$dF_2 = H_3 \neq 0$.  T-dualizing this configuration once would give a
D2-brane on a twisted torus. However, the 2-cycle the D2-brane should
wrap no longer exists in the integer homology of the twisted
torus. Thus, like the original D3-brane, this D2-brane can not exist.
If these branes cannot exist, neither can the D0-brane one would get
by performing two more T-dualities.  This indicates an apparent lack
of the notion of spacetime points on a $T^3$ with $R$-flux. It is in
this sense that we say that the $R$-flux gives a background which is
not even locally geometric.  A further discussion of the properties of
branes in backgrounds with nongeometric fluxes appears in
\cite{Lawrence:2006ma}.

We need some understanding of how to combine these generalized flux
objects in a consistent manner.  These new structures introduce new
terms in the NS-NS Bianchi identities and the R-R tadpoles/Bianchi
identities. The need to satisfy Bianchi identities and tadpole
cancellation conditions constrains the admissible fluxes in a general
compactification.  For the R-R fluxes, the constraints found in \cite{stw} can be written
concisely as \beq
\label{eq:constraints-R}
(\bar H\wedge +f\cdot+Q\cdot+R\contract) \bar \F = 0.
\eeq
Here $\bar \F$ is the formal sum of the integrated R-R field strengths, 
and we have used the notation
\beq
R\contract\omega ^{(p)}\equiv R ^{abc}\omega_{abc
a_4\ldots a_p}
\eeq
in addition to
\beq
f\cdot\omega ^{(p)} = f ^ a_{[bc}\omega_{| a | a_2\ldots a_p]},
\spacer 
Q\cdot\omega ^{(p)} = Q ^{ab}_{[c}\omega_{| ab | a_3\ldots a_p]},
\eeq
contracting all upper indices and antisymmetrizing all 
uncontracted
lower indices.
This is the natural action of a mixed tensor on a $p$-form.  Recalling
that the standard R-R tadpoles/Bianchi identities can be written in
terms of an $H$-twisted differential operator,
\beq
(d+H\wedge)\F \equiv d_H\F = 0,
\eeq
the constraints \er{constraints-R} have a suggestive interpretation 
as the action of an operator defined by acting with all generalized 
NS-NS fluxes,
\footnote{In \er{constraints-R}, there is no term corresponding to the
derivative operator $d$; this is because we have integrated the
differential expression to obtain a topological formula.  The
contraction with $f$ contains the information coming from $d$ when the
forms are expressed in terms of the invariant basis $\eta^a$ on, for
example, a twisted torus.  We expect that some generalization of these
formulas should hold without integration, in which case there should
be additional local contributions from $d$, though it is not clear what this should mean when the compactification space is not a manifold. }
\beq
\cD\equiv H\wedge +f\cdot+Q\cdot+R\contract
\eeq
 on the R-R fluxes $\bar\F$,
\beq
\cD\bar \F = 0.
\eeq  
Moreover, just as the usual Bianchi identity for $H$,
\beq
d H = 0,
\eeq
can be understood as the condition that the twisted differential
$d_H\equiv d+ H\wedge$ be nilpotent, the entire set of NS-NS
constraints, including the contributions from all generalized fluxes,
are equivalent to a nilpotency condition
\beq
\label{eq:constraints-NS}
\cD ^ 2 = (H\wedge +f\cdot+Q\cdot+R\contract) ^ 2 = 0.
\eeq
In addition to the (bilinear) generalized Bianchi identities for the
NS-NS fluxes, \er{constraints-NS} also incorporates the (linear)
``tracelessness'' conditions $f ^ a_{ab} = 0 = Q ^{ab}_a$.

Now that we have rewritten the constraints in a more covariant form as
\er{constraints-R}, \er{constraints-NS}, we have eliminated the
explicit dependence on a particular choice of coordinate system.  This
provides a natural point of departure for considering these
nongeometric structures on manifolds more complicated than the torus,
though we do not pursue this further here.  Discussion of the two-fold
$T $-dual of $H $-flux in the context of Calabi-Yau compactification
can be found in \cite{Benmachiche:2006df}.

While any individual nongeometric flux is T-dual to $H$-flux and has a
reasonably straightforward interpretation as discussed above, a
general combination of geometric and nongeometric fluxes can never be
brought to a description in terms of conventional fluxes in any
duality frame, and may be more difficult to interpret.  Furthermore,
when multiple fluxes are combined, the nature of the compactification
manifold may change dramatically.  For example, including ``geometric
fluxes'' $f ^ a_{bc}$ which give rise to an algebra which is not
nilpotent gives rise to a space in which the fluxes no longer have the
same interpretation as geometric twists.  A simple example of this is
compactification on $S^3$, which produces $f$'s which are the
structure constants of $SU(2)$ \cite{km}.  Performing a T-duality on
the fiber of the Hopf fibration of $S^3$ effectively performs a
T-duality on one of the indices of the $f$'s and gives a space which
can be described geometrically as $S^1 \times S^2$ with one unit of
$H$ flux \cite{bem}.  By formally raising and lowering the indices on
the $f$'s, however, we see that this compactification should be
described by fluxes which we would label as one $H$ and two $Q$'s
\footnote{We would like to thank David Marks for pointing out this
  example to us.}.  While this background is not a solution of the
  equations of motion and is not supersymmetric, it suggests that when
  the algebra is not nilpotent, the question of lifting the background to
  ten dimensions becomes more subtle, and that we should be cautious
  in our assumptions about which backgrounds are geometric and which
  are not.

To be clear, in this paper we will consider a four-dimensional
effective theory with a specific form for the superpotential.  Each
individual term in the superpotential has a clear interpretation in
terms of some individual (generalized) flux on an underlying torus, in
the absence of other fluxes.  However, for the explicit solutions we
find, the gauge algebra of the resulting theory is not nilpotent, and
thus the lift of these solutions to ten dimensions is not
straightforward.  While the individual fluxes appearing in our
superpotential may have nongeometric interpretations, we cannot rule
out the possibility that there may be a completely geometric
description of these vacua in ten dimensions.  For simplicity, we will
continue to use the language of flux compactification on a torus, but
the reader should bear these caveats in mind.

\subsection{A simple model: the symmetric $T ^ 6/\bZ_2 $ orientifold}

In this paper we will carry out the study of generalized flux
compactifications on the simple symmetric torus orientifold $(T ^ 2)
^3/\bZ_2$ which we initiated in \cite{stw}.  As in \cite{stw}, the
underlying compactification we consider is type II on a $T ^ 6$
orientifold where the complex structure is restricted to be diagonal
and symmetric by choosing fluxes which are invariant under cyclic
permutations of the three 2-tori.  Thus, the complex structure
parameters are proportional to the identity matrix, $\tau_{ij}
=\tau\delta_{ij}$. We will consider this orientifold in two T-dual
descriptions, namely as a IIB compactification with O3-planes and as a
IIA compactification with O6-planes. This model has three complex
moduli: $\tau$, which is the complex structure parameter of the torus
in IIB and the K\"{a}hler modulus in IIA; $S$, the axiodilaton; and
$U$, which is the K\"{a}hler modulus in IIB and the complex structure
modulus in IIA.

Allowing all generalized
fluxes which are consistent with the symmetric restriction yields a
polynomial superpotential for the four-dimensional theory:
\begin{eqnarray}
\label{eq:w}
W & = & a_0 - 3a_1 \tau + 3a_2 \tau^2 - a_3 \tau^3\\  
\nonumber
& & \hspace{0.2in} + S (-b_0 + 3b_1 \tau - 3b_2 \tau^2 + b_3 \tau^3)
\\ 
 \nonumber  & &
 \hspace{0.2in} + 3 U (c_0 + (2c_1 - \tilde{c}_1)
 \tau - (2c_2-\tilde{c}_2) \tau^2 -c_3
 \tau^3)\\
&\equiv & P_1 (\tau) +SP_2 (\tau) +UP_3 (\tau).  \nonumber
\end{eqnarray}
The integer coefficients $a_i, b_i, c_i$ correspond to the number of
units of integrated fluxes on cycles of the torus.
\footnote{The flux integer $c_1$ is in the notation of \cite{stw} $c_1
\equiv \hat{c}_1 =\check{c}_1$, and likewise for $c_2$. We will use
the notation $c_i$ to refer to all coefficients in $P_3 (\tau)$.}  The
$a_i$ come from R-R fluxes in both IIA and IIB.  The $b_i$ in IIB come
from $H$-flux on various cycles of the torus, while in IIA they are
given by (in ascending order) $H, f, Q, R$.  The $c_i$ in IIB are all
due to $Q$-flux, and in IIA are again given by (in ascending order)
$H, f, Q, R$. For further details, we refer the reader to \cite{stw}.
In our conventions, the tree-level K\"{a}hler potential is given by
\begin{equation}
K = -3\ln (-i (\tau -\bar{\tau}))
 -\ln (-i (S -\bar{S}))
 -3\ln (-i (U -\bar{U})) \,.
\label{eq:Kaehler}
\end{equation}
Equations \er{w} and \er{Kaehler} determine the scalar potential, which 
has the usual form
\beq
\label{eq:v}
V = e^K \left(\sum_{i,j \in \{\tau, U, S \}} K^{ij} D_i W \overline{D_j
  W} - 3 |W|^2 \right).
\eeq

The R-R tadpole constraints \er{constraints-R} applied to this model 
yield the restrictions on the integer coefficients
\beq
\label{eq:ab}
a_0 b_3-3 a_1 b_2+3 a_2b_1-a_3b_0 = 16
\eeq
and
\beq
\label{eq:ac}
a_0 c_3+a_1 (2c_2-\tilde{c}_2)
-a_2 (2c_1 -\tilde{c}_1)-a_3c_0 = 0.
\eeq
The nonvanishing right hand side of \er{ab} is the contribution from
the orientifold planes.  The NS-NS Bianchi identities and tadpoles, in
this model, give a set of constraints relating the $b_i$ and the
$c_i$:
\begin{eqnarray}
\label{eq:bc1}
c_0 b_2-\td{c}_1 b_1+c_1 b_1-c_2 b_0 & = & 0\\
c_1 b_3-c_2 b_2+\td{c}_2 b_2-c_3 b_1 & = & 0
\label{eq:bc2}\\
c_0 b_3-\td{c}_1 b_2+ c_1 b_2-c_2 b_1 & = & 0
\label{eq:bc3}\\
\label{eq:bc4}
c_1 b_2-c_2 b_1+\td{c}_2 b_1-c_3 b_0 & = & 0,
\end{eqnarray}
as well as a set of constraints among the $c_i$:
\begin{eqnarray}
c_0\td{c}_2-c_1 ^ 2+\td{c}_1c_1-c_2 c_0 & = & 0
\label{eq:cc1}\\
c_3\td{c}_1-c_2 ^ 2 +\td{c}_2c_2-c_1 c_3 & = & 0\\
c_3 c_0-c_2c_1  & = & 0
\label{eq:cc3}.
\end{eqnarray}
Of the seven NS-NS constraints (\ref{eq:bc1}--\ref{eq:cc3}), four are
independent.  Including the R-R constraints (\ref{eq:ab}--\ref{eq:ac})
then gives six independent constraints, reducing the number of
independent flux integers from fourteen to eight.

Finally, in order to avoid additional subtleties, we impose the
further restriction that all flux coefficients should be even.  This
restriction is motivated by the observation that even numbers of flux
quanta on an orientifolded manifold lift to integer quanta on the
original manifold, while odd numbers of flux quanta require a more
sophisticated construction incorporating exotic orientifold planes \cite{Frey:2002hf}.
For simplicity, we will thus consider only even numbers of flux quanta
for all of the generalized fluxes we consider here.  For any
individual nongeometric NS-NS flux, this restriction to even integers
only can be understood as the T-dual description of requiring the
number of units of $H$-flux to be even.  While this argument from
T-duality is insufficient to fully describe a general combination of
NS-NS generalized fluxes, where the new fluxes can never be T-dualized
completely to $H$-flux, it is a strong motivation to consider only
even flux coefficients.

\section{Equations of Motion}
\label{sec:eom}

Let us now discuss solutions to the tree level equations of motion in
the effective theory defined by the superpotential (\ref{eq:w}) and
the K\"{a}hler potential (\ref{eq:Kaehler}).  The scalar potential
comes entirely from $F$-terms, as in this model the remaining fields are not charged under the gauge group, and there is therefore no $D$-term contribution.  Thus, to find
supersymmetric solutions, it suffices to solve the $F$-flat conditions
$D_\alpha W = \partial_\alpha W+ (\partial_\alpha K) W = 0$.

In Subsection \ref{sec:equations} we summarize the equations we are
interested in solving.  In \ref{sec:special} we describe some simple
special cases of solutions.  In \ref{sec:valid} we discuss the
criteria for validity of solutions.

\subsection{Equations}
\label{sec:equations}

The equations of motion we are interested in solving are the $F$-flat
conditions
\begin{eqnarray}
P_1 (\tau) + \bar{S}P_2 (\tau) +UP_3 (\tau)& = &  0 \label{eq:es}\\
P_1 (\tau) + S P_2 (\tau) + \left({2\over 3} U+{1\over 3}\bar{U} \right) P_3 (\tau) & = &  0  \label{eq:eu}\,\\
(\tau -\bar{\tau})  \partial_\tau W -3W & = & 0  \label{eq:et}\,.
\end{eqnarray}
Using equations \er{es} and \er{eu}, one finds that physical solutions 
must satisfy
\beq
\label{eq:tau1}
\im (P_2\bar P_3) = 0.
\eeq
One may readily solve for $U$ and $S$ as functions of $\tau$, leaving
\er{tau1} and a twelfth-order (real) polynomial equation for the real
and imaginary parts of $\tau$ coming from \er{et}.  Thus, we must
simultaneously solve this twelfth-order polynomial and a (real)
fifth-order polynomial from \er{tau1}, which in general must be done
numerically.  Once we have a solution $\tau_0, S_0, U_0$, we can read
off the string coupling,
\begin{equation}
g ={1\over\Im S_0},
\label{eq:g}
\end{equation}
and the cosmological constant
\beq
\Lambda =V(\tau_0, S_0, U_0) =-3 e ^{K (\tau_0, S_0, U_0)} | W (\tau_0, S_0, U_0)| ^ 2.
\eeq

We consider physical solutions to equations (\ref{eq:es}--\ref{eq:et})
to be only those with positive values for the imaginary parts of all
moduli.  In the absence of fluxes, this requirement is straightforward
to interpret, as the imaginary parts of the moduli control the string
coupling and geometric information about the internal manifold, and
must be strictly greater than zero.  We assume that this positivity
requirement continues to hold in the presence of arbitrary
combinations of generalized fluxes, as turning on fluxes simply turns
on new couplings in the scalar sector of the four-dimensional theory.

Relatedly, the real (axionic) parts of the moduli are in our
conventions periodic with unit period.  This periodicity is easily
derived in the underlying toroidal compactification in the language of
either IIA or IIB.  The addition of fluxes, conventional and
nongeometric, again only adds couplings between the scalar fields and
does not affect the range in which they take values.

Our interest here is to systematically study the solutions to the
equations of motion (\ref{eq:es}--\ref{eq:et}) for general choices of
the integer coefficients $a_i, b_i, c_i$ in (\ref{eq:w}).  Not all
choices of flux integers admit a physical supersymmetric solution.
The choices of flux integers that do yield physical solutions
generically fix all moduli and are generically not gauge equivalent to
any vacua which have a conventional geometric interpretation in either
type IIA or IIB.  We will discuss two particularly well-behaved
families of such vacua in Section 4, and describe the general
solutions in Section 5.

In order to understand the solution space of this model, we must
understand how to avoid multiply counting solutions which are related
by the modular group of the compactification.  In our simple torus
model, the non-compact portion of the modular group is given by the
integral shift symmetries of the three axionic scalar fields.  The
shift $\tau\rightarrow\tau +n$ combines with the inversion
$\tau\rightarrow-1/\tau$ to give a factor of $SL (2, \bZ)$.  The
S-duality transformation $S \rightarrow -1/S$, which would combine
with the shifts $S\rightarrow S+m$ to give another factor of $SL (2,
\bZ)$, maps generalized NS-NS fluxes to additional nongeometric
structures which mix the metric with R-R fields \cite{stw, acfi}.  As
S-duality therefore takes us out of the class of compactifications we
consider here, it is not a factor in the modular group and we will not
study it further here.  The same is true about the inversion
transformation $U\rightarrow -1/U$.  Each of these transformations of
the moduli is accompanied by a corresponding transformation of the
fluxes to leave the equations of motion invariant.  We give the
explicit form of all of these transformations in
appendix~\ref{a:modular}.  These transformations descend from large
gauge transformations, and therefore configurations related by these
transformations are physically equivalent. In addition, there is a
$\bZ_2$ factor in the modular group which flips the signs of all
fluxes, taking $W \rightarrow - W$; this transformation does not
produce a physically distinct vacuum and as such is also a gauge
transformation.

In addition to the modular group discussed above, there are a number
of additional transformations relating vacua which, despite having
identical spectra, are physically inequivalent. First, there is a sign
flip taking $\tau, S, U \rightarrow -\bar \tau, - \bar S, - \bar U$,
with an accompanying transformation on the fluxes leaving the
equations of motion invariant. Additionally, one may shift the axions
by rational numbers, e.g. $U \rightarrow U + 1/n$ or $S \rightarrow S
+ 1/n$, which is possible whenever the resulting fluxes are still even
integers.  Both the sign flip and the fractional shift transformations
relate vacua with physically distinct expectation values for the
axions but identical spectra of particle masses, coupling constant,
and cosmological constant. Thus, we have
relations between different four-dimensional theories since the
transformations act on the couplings as well as the fields. One can
therefore think of theories related by these transformations as being part
of a discrete moduli space of a larger theory.

\subsection{Some special cases}
\label{sec:special}

There are several classes of possible flux configurations which are of
special interest: fluxes which have an interpretation as a geometric
IIB compactification, fluxes which have an interpretation as a
geometric IIA compactification, and fluxes yielding solutions which
have a vanishing expectation value for the superpotential, giving a
Minkowski vacuum at tree level.  Each of these subclasses of flux
integers seems to be a set of measure zero with respect to the full
set of possible flux configurations admitting physical solutions.  We
will discuss these cases in this section, and treat more
general cases in the remainder of the paper.

\subsubsection{Geometric IIB vacua}

If one takes all the $c_ i$ to vanish, then the resulting
compactification is simply a standard IIB flux compactification on the
torus, and the superpotential \er{w} reduces to the well-known
Gukov-Vafa-Witten superpotential \cite{gvw}.  The resulting
vacua on the torus have been studied in \cite{dgkt1,kst,iib}. As is
well-known, these fluxes are not sufficient to stabilize all moduli at
tree level.  While in principle these configurations would appear as a
subset of the compactifications we consider here, our main interest in
this paper is in vacua with all moduli stabilized, so we will not
discuss geometric IIB vacua further here.  In particular, these vacua
will not be included in the sets of vacua studied in
Section~\ref{sec:statistics}.

\subsubsection{Geometric IIA vacua}

If the fluxes $b_2, b_3$ and $ c_2,\tc_2, c_3$ all vanish, then all
the remaining terms in the superpotential have a geometric
interpretation in IIA; such toroidal compactifications with geometric
flux have been studied in \cite{hre,Kachru:2002sk,dkpz,vz1, cfi}.  On
the symmetric torus, however, there is an additional interplay between
the equations of motion and the constraints which drastically limits
the kinds of supersymmetric vacua one may obtain in this way.

Since for a geometric solution in IIA, $P_2 $ and $P_3$ are linear, 
equation \er{tau1} for $\tau$ reduces to
\beq
(\im\tau )(3 b_1 c_0+b_0 (2c_1-\tc_1)) = 0.
\eeq
Requiring $\im\tau >0$ then yields a constraint on the fluxes
\beq
\label{eq:first}
3 b_1 c_0+b_0 (2c_1-\tc_1) = 0.
\eeq
Consider, however, the tadpole constraints for this set of fluxes:
\barray
\label{eq:ab2}
3 a_2 b_1-a_3 b_0 & = & 16\\
\label{eq:ac2}
a_2 (2c_1-\tc_1) +a_3 c_0 & = & 0\\
(c_1-\tc_1) b_1 & = & 0\\
c_1(c_1-\tc_1)  & = & 0.
\earray
Let us suppose first that neither $b_1$ nor $(2c_1-\tc_1) $ is zero.  
Then using \er{first}, we can trade the $c_i$ in \er{ac2} for $b_i$ to 
find
\beq
3 a_2 b_1-a_3 b_0 = 0,
\eeq
which is, of course, inconsistent with \er{ab2}.

The only way to avoid this inconsistency is to take either $b_1$ or
$(2c_1-\tc_1) $ to be 0.  Then the constraints enforce $c_1 =\tc_1 =
c_0 = 0$, for both $b_1 = 0$ and $b_1\neq 0$.  In either case, this is
simply an alternate description of a conventional IIB
compactification.

If one chooses to cancel the tadpole from the orientifold planes using
D-branes instead of fluxes, then one may set the right hand side of
\er{ab2} to zero.  The constraints then allow for a wider range of
possible solutions, including those with no geometric interpretation
in IIB.  We will limit our analysis here, however, to
the tadpole as given in \er{ab2}, without D-branes.

\subsubsection{Vacua with $W = 0$}

From the equations of motion, one may show that vacua where the
superpotential vanishes at the minimum are obtained only when
\beq
P_1 (\tau_0 ) =P_2 (\tau_0 ) =P_3 (\tau_0 ) = 0,
\eeq
where $\tau_0$ is the value of $\tau$ which solves the $F$-flat
equations.  It is then clear, as we need $\im\tau >0$, that a
necessary condition for a $W = 0$ solution is that the polynomials $P_
i$ all share a common quadratic factor, $P$, which vanishes at
$\tau_0$.  The remaining nontrivial equations of motion then fix two
of the four remaining degrees of freedom. While Minkowski vacua are
very interesting, again, we are focusing our attention on vacua with
all moduli stabilized, and therefore will not examine $W=0$ solutions
further. For a novel approach to this class of solutions, see
\cite{Gray:2006gn}.

\subsection{Validity of solutions}
\label{sec:valid}

We are interested in supersymmetric solutions of the
four-dimensional effective theory defined by the superpotential \er{w}
and the K\"{a}hler potential \er{Kaehler}.  
In order for these solutions to be at all trustworthy, we need to have
a small value for the string coupling $g$, given through (\ref{eq:g}).
Otherwise, nonperturbative string effects may radically change the
structure of the low-energy theory.

The next critical question in effective field theory calculations in a
typical flux compactification is whether the moduli are stabilized at
masses light enough to render the computation reliable when
perturbative string effects are included -- that is, whether the
effective field theory describes effects at energy scales which are
controllably smaller than the energy scales of higher Kaluza-Klein
modes, winding modes, and excited string modes.  The analogous
question for a compactification incorporating nongeometric twists is
more subtle as in the absence of a well-defined critical sigma model
we do not know precisely how to estimate the mass scales of the string
modes (in general the distinction between Kaluza-Klein and other
inherently stringy modes such as winding modes will break down).
Given some such concrete stringy solution, we might hope in principle
to find a rough estimate of the scale of the string modes to compare
with the masses of the modes retained in the four-dimensional
effective theory.

We have found sets of supersymmetric solutions to the effective field
theory which have tunably small string coupling, cosmological
constant, and masses for the moduli, as we will detail in the
following section.  These solutions have some very interesting
properties.  As the string coupling can be made arbitrarily small,
they are well in the perturbative regime.  The mass scales computed in
the effective theory can also be made parametrically small, although,
again, we do not have a reliable method of computing the scale of the
effective UV cutoff in our models for comparison.  Here we take, as a
zeroth-order estimate, the scale of heavier modes to be set by those
scales which, in the absence of flux, would set the masses of momentum
and winding modes on the torus.  The mass scales in our families of
solutions can indeed be made parametrically lighter than this estimate
for the cutoff.  Our solutions do contain non-nilpotent algebras,
however, (as do \cite{dkpz, vz1}) and thus are in the class of vacua
which seem most difficult to lift to ten dimensions.  By analogy with
the geometric cases discussed in \cite{km}, we might be concerned that
our na\"ive method of estimating the cutoff scale, based on an
underlying torus geometry, may be inaccurate, so that some of the
moduli masses may be comparable to the cutoff scale.  In this case,
our vacua would need to be augmented with other modes with comparable
masses to the fields already included in the effective
four-dimensional theory in order to construct a full-blown string
model.  On the other hand, the existence of supersymmetric vacua with
parametrically controlled string coupling, moduli masses, and
cosmological constant suggests strongly that these vacua can be lifted
in some way to string theory.

\section{Families of Nongeometric Vacua}
\label{sec:families}

In this section we discuss two particularly interesting infinite
families of solutions to the equations of motion presented
in the previous section.  These infinite families precisely saturate
the tadpole (which as discussed in Section \ref{sec:review} is equal
to 16) without the inclusion of additional D-brane sources.

The fluxes which lead to our interesting infinite families of
solutions are all modular equivalent to fluxes having
\beq 
P_2(\tau) = b (\tau-1)^3 
\eeq
for some integer $b$, i.e. all the $b_i$ are equal.  This form for the
$b_i$ imposes restrictive constraints on the $c_i$.  If in addition
$c_3=0$, there are only two possible choices for the remaining $c_i $
which satisfy the constraints (\ref{eq:bc1}--\ref{eq:cc3}).  These two
options are, in the notation $(c_0, (c_1,\tc_1), (c_2,\tc_2), c_3) $,
\begin{eqnarray}
\label{eq:fam1}
\mathrm{either}\,\,  & & (2(n+m), (-2 m, 2 n), (0, 2 m), 0)\\ 
\label{eq:fam2}
\mathrm{or}\,\, & & (2 (n+m), (0, 2 m), (2 n, 2 n), 0). 
\end{eqnarray}
Each of these choices for the $c_i$ leads to an interesting set of
physical solutions, as we will see.  Given these choices for the $b_i$
and the $c_i$, one must then choose $a_i$ which satisfy the RR
constraints (\ref{eq:ab}--\ref{eq:ac}).  A particularly simple way to
solve the RR constraints is to take $a_0 = 16/b$, with the remaining
$a_i = 0$. Since $a_0$ and $b$ must be (even) integers, the only
possible choices for $b$ are 2, 4, and 8. There are other choices one
may make for the $a_i$; we merely choose this one for simplicity and
because we can exactly solve for the moduli.

The first family of solutions we consider is (\ref{eq:fam1}), with the
$a_i$ chosen as in the previous paragraph. Using the notation $(a_0,
a_1, a_2, a_3),\, (b_0, b_1, b_2, b_3),\, (c_0, (c_1,\tc_1),
(c_2,\tc_2), c_3)$, the fluxes here are
\begin{equation}
(16/b,0,0,0),\; (b,b,b,b),\; (2(n+m),(-2m,2n),(0,2m),0).
\label{eq:fullfam1}
\end{equation}
These fluxes yield physical solutions for the moduli
\begin{eqnarray}
\label{eq:t0}
\tau_0 &=& \left (  \frac{n+m}{m} \right ) \pm \left ( \frac{n}{m} \right ) i,   \\
S_0 &=& \frac{m^3}{b^2n^3}\left (-8 \pm 2i \right ),  \\
\label{eq:u0}
U_0 &=& \frac{m}{bn^2} (4\pm2i).
\end{eqnarray}
The $\pm$ in the imaginary parts should be picked to make the
imaginary parts positive, i.e. $+$ when $m/n>0$ and $-$ when $m/n <
0$. The same sign must be picked in all three moduli to ensure a
solution to the equations of motion.  This family of solutions has
string coupling 
\beq 
g = \left | \frac{b^2n^3}{2m^3} \right | 
\eeq 
and
cosmological constant 
\beq \Lambda = - \left | \frac{3b^3n^6}{16m^3}
\right | .  
\eeq 
This family is thus easy to tune to small string
coupling and cosmological constant by fixing $n$ and taking $m$ to be
large. Notice that it is impossible to tune both $g $ and $\Lambda $
to be finite; there must be an accumulation point for one or both
quantities at either zero or infinity in any infinite set.

Meanwhile, the fields $\tau, S, U $ have a mass matrix given by 
\beq
M_{ij} =\left(\begin{array}{cccccc}
 \pm\frac{21b^3n^4}{16m}&0&0&\frac{3b^5n^8}{32m^5}&0&\frac{-3b^4n^7}{32m^3}\\
 0&\pm\frac{33b^3n^4}{16m}&\frac{-3b^5n^8}{16m^5}&0&0&0\\
 0&\frac{-3b^5n^8}{16m^5}&\pm\frac{b^7n^{12}}{32m^9}&0&\pm\frac{3b^6n^{11}}{64m^7}&0\\
 \frac{3b^5n^8}{32m^5}&0&0&\pm\frac{b^7n^{12}}{32m^9}&0&\pm\frac{3b^6n^{11}}{32m^7}\\
 0&0&\pm\frac{3b^6n^{11}}{64m^7}&0&\pm\frac{3b^5n^{10}}{32m^5}&0\\
 \frac{-3b^4n^7}{32m^3}&0&0&\pm\frac{3b^6n^{11}}{32m^7}&0&\pm\frac{3b^5n^{10}}{16m^5}
\end{array}\right),
\label{eq:massmatrix}
\eeq
in the basis $\{\re \tau,\im \tau,\re S, \im S, \re U,\im U \} $. The
choices of sign $\pm $ in the matrix elements are determined by the
choices of sign in the moduli, (\ref{eq:t0}--\ref{eq:u0}). The
eigenvalues of $M_{ij} $ come in pairs, scaling as $m ^ {-1},\, m ^{-
5}, $ and $m ^{-9} $ in the controllable limit where
$m\rightarrow\infinity $ with $n $ fixed.  Thus all the fields in this
family of solutions have parametrically light masses. Our zeroth-order
estimate for the cutoff scale, by contrast, scales as $m ^ 0$.

The second two-parameter family, with the $c_i$ as given in
(\ref{eq:fam2}), is similar to the one above. The fluxes are
\begin{equation}
(16/b,0,0,0),\; (b,b,b,b),\; (2(n+m),(0,2n),(2m,2m),0),
\label{eq:fullfam2}
\end{equation}
and it is easy to check that the moduli
\begin{eqnarray}
\tau_0 &=& -\left (  \frac{n+m}{m} \right ) \pm \left ( \frac{n+2m}{m} \right ) i ,  \\
S_0 &=& \frac{1}{b^2}\left ( \frac{m}{2m+n} \right )^3 \left ( 8 \pm 2i \right ),\\
U_0 &=& \frac{m}{b(2m+n)^2} (-4\pm 2i)
\end{eqnarray}
solve the $F $-flat equations. As above, the $\pm$ in the imaginary
parts of these moduli should be picked to keep the imaginary parts
positive; the plus sign should be used when $m>0,\, n>-2m$ and the minus
sign should be used when $m<0,\, n>-2m$. Again, the same sign must be
picked in all three moduli to give a solution to the equations of
motion.  These values for the moduli give string coupling
\beq
g =  \left | \frac{b^2(2m+n)^3}{2m^3}  \right |
\eeq
and cosmological constant
\beq
\Lambda = -\left | \frac{3b^3(2m+n)^6}{16m^3} \right | .
\eeq
The masses for the moduli in these vacua are again parametrically light.
These two families have identical spectra but are not equivalent under modular
transformations. 

There are more general choices one can make for the $a_i$ than in the
two families discussed explicitly above. For example, any modular
transformation $S \rightarrow S+q $ or $U \rightarrow U+q $ will take
the fluxes of \er{fullfam1}, \er{fullfam2} to flux sets with more
$a_i$ nonzero.

It is particularly interesting to consider the effect of fractional
shifts of the form $U \rightarrow U + (1/q)$, on these families.
These shifts are permissible whenever $c_i/q$ is an even integer for
all $c_i $. Every time we can perform such a shift, we get a
physically inequivalent theory with identical spectrum.  Each
fractional shift yields another infinite family of vacua, with
different values for the flux parameters $a_i $ and for the
expectation value of $\re U $.  Clearly, the number of such shifts we
can perform is controlled by the factors of the $c_i$ fluxes. But
because the $c_i $ appear only homogenously in the constraint
equations, we are free to scale them up to arbitrarily large
values. As we scale up $m$ and $n$, we will encounter numbers that
have an increasingly large number of factors, and be able to perform
more such shifts. Thus, as $m$ and $n$ increase, the degeneracy of
these vacua increases, becoming infinite in the $m,n\rightarrow
\infty$ limit. 
Degeneracies due to fractional shifts of
the axions are in fact a general, though little-remarked, feature of
flux compactifications; the novelty in the present case is that the
tadpole cancellation conditions fail to prevent arbitrarily large
degeneracies.

In this section we have derived infinite families of solutions to the
four-dimensional field theory associated with general fluxes.  As we have discussed,
it is not clear that we can take even those solutions with small $g $ 
at face value as solutions of a complete string theory compactification.
It is furthermore entirely possible that even if these solutions are
valid for small fluxes, they may break down in the limit of infinite
flux.  Although the average energy density in the internal space due
to the fluxes, given by the cosmological constant, can remain small,
as the fluxes become infinite one might expect that there are regions
where the local energy density becomes large enough to necessitate an
accurate treatment of back reaction.

\section{Statistics of Generic Vacua}
\label{sec:statistics}

So far we have described some special classes of solutions to the
equations of motion (\ref{eq:es}, \ref{eq:eu}, \ref{eq:et}).  To
investigate the properties of general solutions, we proceed
numerically.  Numerical computation of solutions for flux vacuum
equations of this type requires three ingredients.  First, we must
generate a set of (even) fluxes satisfying the constraint equations
(\ref{eq:ab}-\ref{eq:cc3}).  Second, we must numerically solve the
equations (\ref{eq:es}, \ref{eq:eu}, \ref{eq:et}) for the moduli
$\tau, S, U$.  Finally, we must impose a gauge-fixing of the modular
symmetries (\ref{eq:modular-begin}--\ref{eq:modular-end}) so that we
do not count the same ``vacuum'' twice.

In previous analogous work on the statistics of IIB flux
compactifications \cite{dgkt1,stats,ad} only a finite number of
solutions were compatible with a fixed tadpole bound $L$.  Thus, $L$
acted as a natural cutoff, and it was natural to investigate the
growth of the number of solutions as a function of $L$.  In the case
we are investigating here, as we have seen in the previous section,
there are infinite families of solutions even at fixed tadpole $L =
16$.  To proceed numerically we must impose an artificial cutoff on
the set of allowed fluxes.  We then need to compute quantities which
approach well-defined limits as the cutoff is taken to infinity.

Generating fluxes satisfying the constraints is straightforward.
Fixing an upper bound $N$ for the fluxes, so that
\begin{equation}
| a_i |, | b_i |, | c_i | \leq N \,,
\label{eq:flux-region}
\end{equation}
we can scan through an independent subset of the fluxes and solve for
the remaining dependent fluxes using some of the constraints in time
polynomial in $N$.  In general, any particular algebraic solution to the
constraint equations is only valid if certain quantities are
nonvanishing.  For example, solving (\ref{eq:ab}) for $a_0$ is only
possible if $b_3 \neq 0$.  We have generated fluxes using several
different subsets of the equations, and have confirmed that the fluxes
missed in this process comprise an increasingly small fraction of the
set of allowed fluxes as $N$
is increased.

Numerically solving the equations (\ref{eq:es}-\ref{eq:et}) is also
straightforward in principle.  Again, however, any specific approach
to solving the equations assumes certain quantities to be
nonvanishing.  We have carried out a systematic search for solutions
to the $F$-flat equations which stabilize all moduli at tree
level for fluxes within the region (\ref{eq:flux-region}).  The
stabilization of all moduli is generic, although as we have mentioned
in Section \ref{sec:special} there are some cases of special interest
which do not fall into this category.

As discussed in, for example, \cite{dgkt1,ad}, there are two ways of
imposing gauge-fixing.  On the one hand, we can generate all possible
fluxes satisfying the constraints (up to our cutoff), numerically
solve the vacuum equations for all fluxes, and then select only those
solutions which live in a fixed modular region in $\tau,S, U$ space.
On the other hand, we can perform the gauge-fixing at the level of the
fluxes, choosing only fluxes which satisfy a particular gauge-fixing
condition.  In general, if we can gauge-fix at the level of the
fluxes, such that we can efficiently generate only fluxes in the given
modular region, our search will be much more efficient, since
otherwise most of our computer time is spent scanning regions outside
the modular domain; this problem becomes worse as the scale of the
fluxes increases and more copies of the modular domain are probed.

The challenge in gauge-fixing at the level of the fluxes is that there
is not always a simple gauge-fixing choice for the fluxes.  In the
problem we are interested in here, it is easy to fix the modular
transformations $S \rightarrow S + n,\, U \rightarrow U + m$ by
imposing conditions on the fluxes.  It is less straightforward,
however, to fix the $SL(2,\bZ)$ symmetry on $\tau$ at the level of the
fluxes.  We have used the following approach to gauge-fix this
symmetry: when $P_2 (\tau)$ has a pair of complex roots, we can use it
to determine a reference value $\tau_r$ which is the root of $P_2$ in
the upper-half complex plane. We then gauge-fix the $SL(2,\bZ)$
symmetry by requiring $\tau_r$ to be in the standard fundamental
domain ${\cal F}=\{ \tau : | \tau| \geq 1, -1/2 < \Re \tau \leq 1/2, | \tau | = 1 \Rightarrow \Re \tau \geq 0 \}$.  When $P_2 (\tau)$ has all real roots, but $P_3
(\tau)$ has a pair of complex roots, we choose $\tau_r$ to be the
complex root of $P_3$ with positive imaginary part.  We then again fix
the $SL(2,\bZ)$ symmetry by taking $\tau_r\in {\cal F}$.  The
complexity of the roots of a cubic polynomial depends on a
discriminant inequality on the coefficients, much as for a quadratic
polynomial.  Thus, the subset of fluxes which cannot be gauge-fixed in
this way represents a fraction of order $1$ of the full set of fluxes.

We have generated sets of fluxes up to $N = 30$ which satisfy the
following explicit modular and sign fixing conditions
\begin{itemize}
\item $\tau_r \in{\cal F}$
\item  $c_3 \geq 0$, if $c_3 = 0$ then $b_3 \geq 0$
\item $b_0 \geq 0$, if  $b_0 = 0$ then $c_0 \geq 0$
\item $(a_0, a_3) = \alpha (b_0, b_3) + 3 \beta (c_0, c_3),$ where
 $0 \leq \alpha, \beta <1$
\end{itemize}
Restricting to these fluxes completely fixes the modular freedom,
though again some sets of measure zero are lost.  Here we are fixing two
choices of sign, both the choice of overall sign for the fluxes, which
relates two different descriptions of the same vacuum, as well as the
global transformation \er{global-flip}, which relates two inequivalent
but degenerate vacua.  Therefore for each solution we find, there is
an additional solution related by the transformation
\er{global-flip}. We find that the number of gauge-fixed fluxes in the
region (\ref{eq:flux-region}) scales roughly as $N^4$.

We have explicitly solved the equations of motion
(\ref{eq:es}-\ref{eq:et}) for this set of gauge-fixed fluxes.  We find
that roughly 20\% of all fluxes yield physical supersymmetric
solutions.  This fraction did not change significantly as $N$
increased.  A plot of the distribution of string couplings $g$ and
cosmological constants $\Lambda$ is shown in Figure~\ref{f:gl1}, for
the data up to $N = 20$.  
\FIGURE{ \epsfig{file=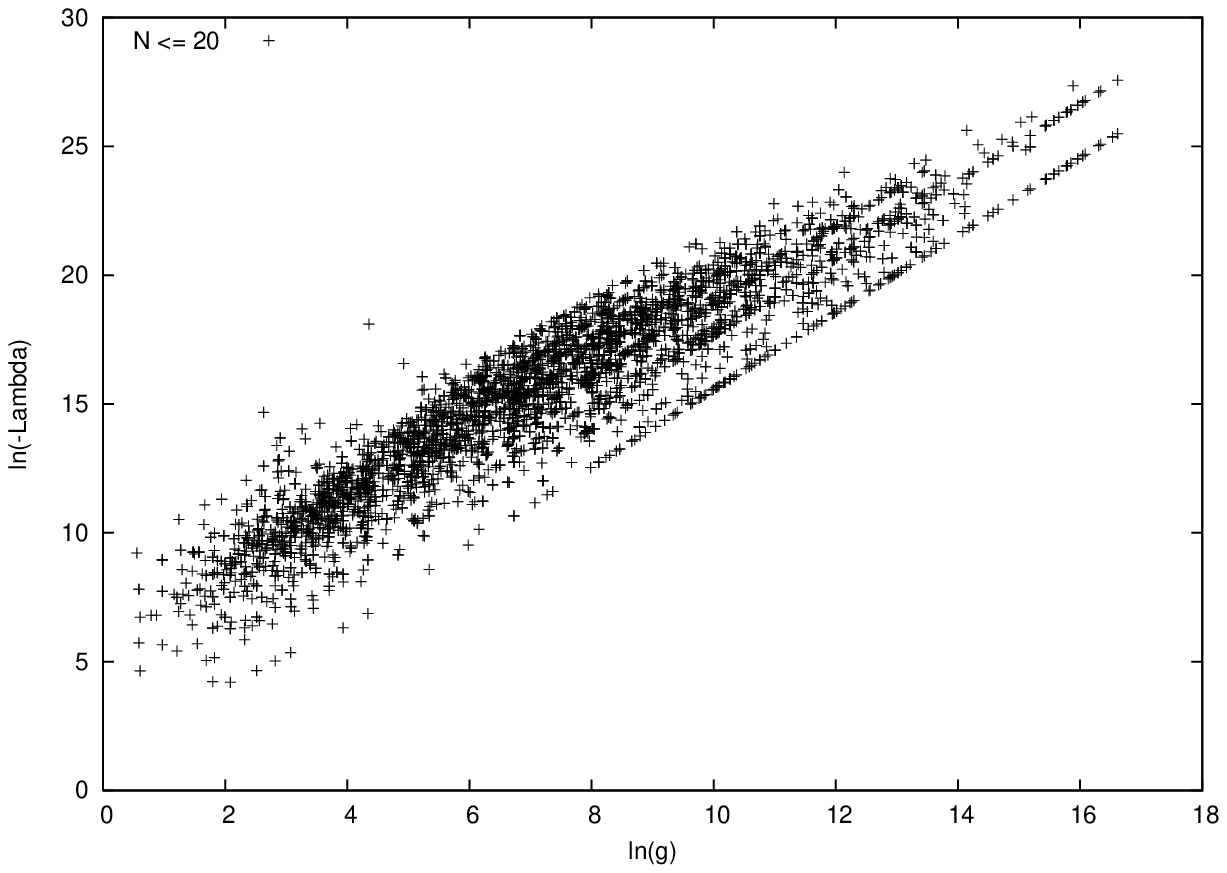,width=15cm}
\caption{\footnotesize Log-log plot of $g, \Lambda$ for gauge-fixed
  fluxes up to $N = 20$}
\label{f:gl1}
} As can be seen from the figure, large string coupling is correlated
with large $\Lambda$.  This is not surprising, since $g = 1/{\rm Im\;}
S_0 $ appears as a factor of $e^K$.  In addition one may see from
studying the equations of motion that $g $ does not scale with $P_ 3
$, while $\Lambda $ scales as $P_3 ^ 3$, so that rescaling the $c_i $
increases the cosmological constant and correspondingly decreases $U$
without affecting $g$.
This demonstrates that the distribution shown in Figure~\ref{f:gl1} will have a tail reaching
up to $-\Lambda \rightarrow \infty$ at fixed $g$ as $N \rightarrow \infty$.

What is perhaps even more interesting is that the overall shape of the
distribution shown in Figure~\ref{f:gl1} does not change appreciably
as $N$ increases.  As far as we can tell from our sample, the
distribution of vacua in $g$-$\Lambda$ space is independent of the
scale $N$ of the fluxes.  This implies in particular that unless there
is some dramatic qualitative change at much larger $N$, there are an
infinite number of solutions of the $F $-flat equations in fixed
finite regions of $g$-$\Lambda$ parameter space.  Moreover, the vacuum
expectation values for the non-compact scalar fields $\im \tau $ and
$\im U $ also accumulate according to a distribution whose shape is apparently independent of $N$.  This leads to an infinite accumulation of solutions in fixed finite regions of parameter space.
\FIGURE{ \epsfig{file= 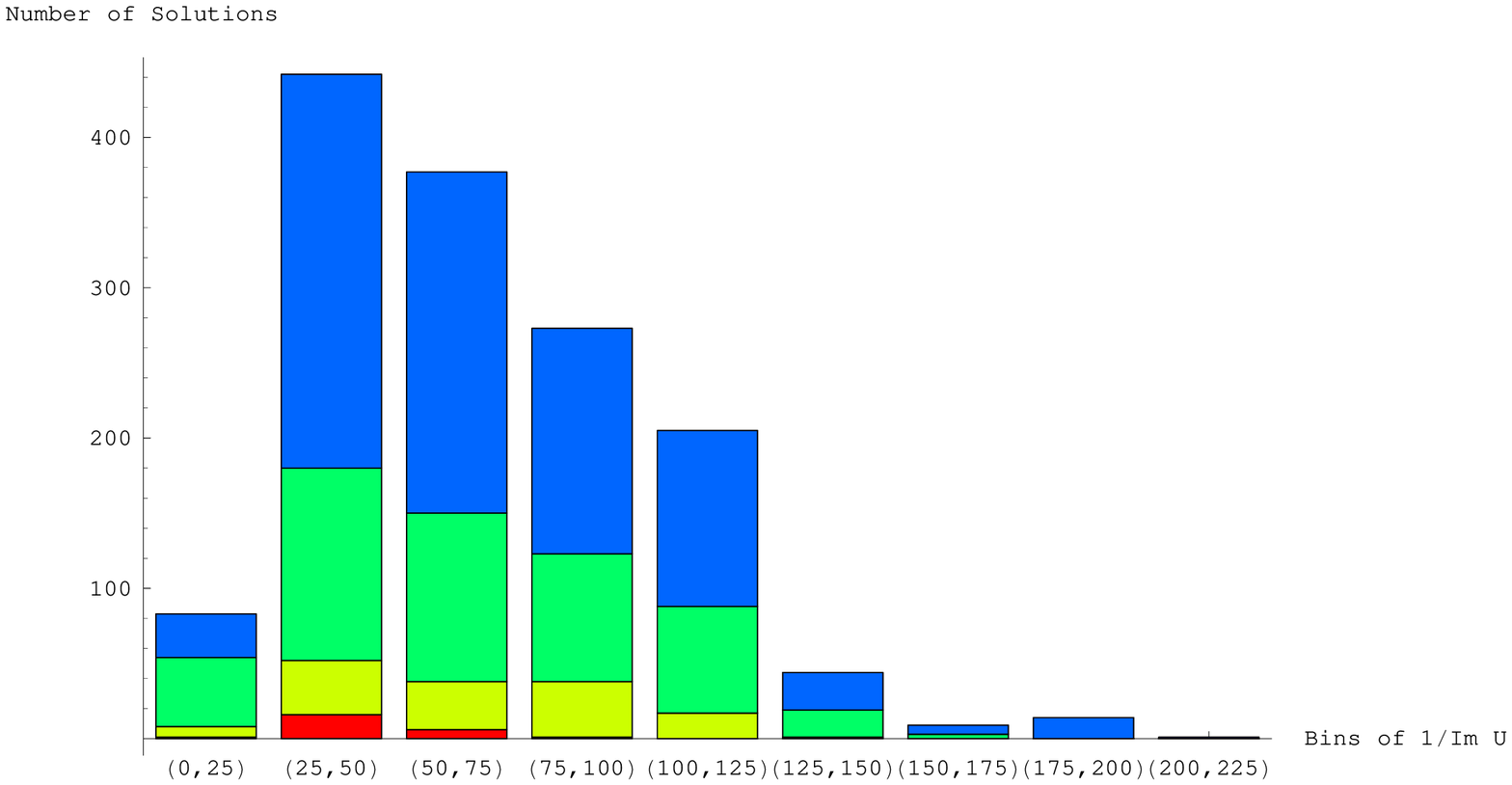,width=15cm}
\caption{\footnotesize Distribution of values of $1/\im U $ as a
  function of maximum flux $N $ for solutions within a particular
  range of $g $ and $\Lambda $.
Successive shaded regions correspond to $N = 8, 12, 16, 20$ respectively.}
\label{f:u}
} In Figures~\ref{f:u} and \ref{f:t} we select one bounded region in $g$-$\Lambda$
space and show how the number of solutions within given ranges of $\im
\tau $ and $\im U $ increases with the maximum flux $N $.  
The region in $g$-$\Lambda$
space we have chosen to illustrate in the figures is $10 <\ln \,(-\Lambda) <15$,
$1 <\ln g <8 $.  In Figure~\ref{f:u} we show the distribution of $\im U $ 
for solutions which lie within this bounded region in $g$-$\Lambda$ space.  
In Figure~\ref{f:t}, we show the distribution of $\im\tau $ for the same set of solutions.
\FIGURE{
\epsfig{file= 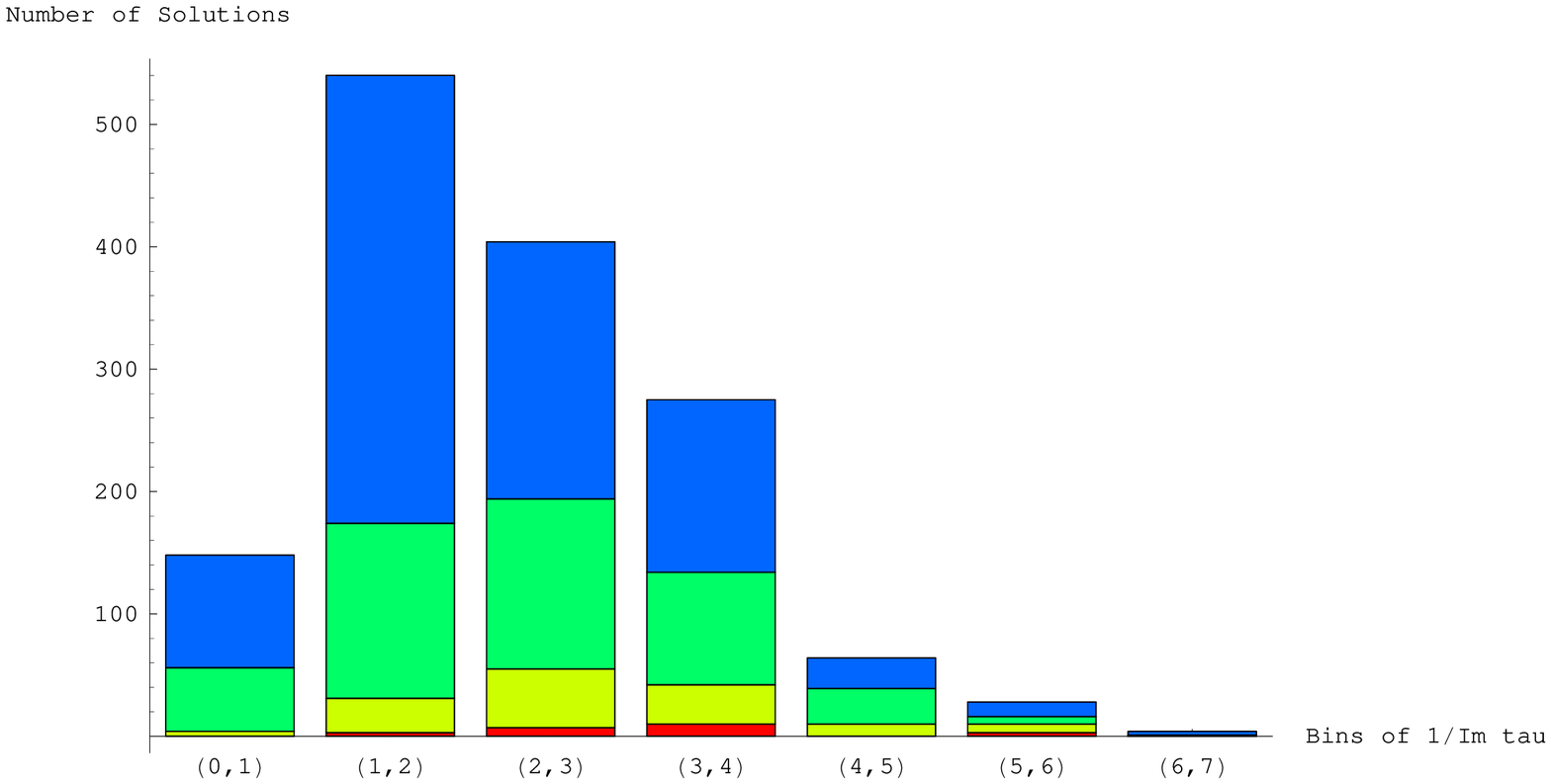,width=15cm}
\caption{\footnotesize Distribution of values of $1/\im\tau $ as a
  function of maximum flux $N $ for solutions within a particular
  range of $g $ and $\Lambda $.
Successive shaded regions correspond to $N = 8, 12, 16, 20$ respectively.}
\label{f:t}
} The data have been binned into ranges of useful size for both
variables.  The gradations of color/grayscale in the graphs show how the
solutions accumulate as a function of maximum flux size $N $; we have
plotted the data for $N = 8, 12, 16, $ and $20$, with a separate color/shade
for each successive value of $N$. 
As is evident in both figures, the distribution of vacua appears
to be fairly independent of the scale $N$ of the fluxes, and exhibits no strong
indication of running to the boundaries of moduli space.  While these
``vacua" are completely untrustworthy, lying at large string coupling,
this illustrates that for this model there seem to be
an infinite number of distinct
tree-level SUSY solutions for a family of gauged supergravity theories in a finite volume
of moduli space.  
This is the only example we know of where even in the tree-level
low-energy approximation there is  such an infinite family of vacua.
If these vacua correspond to good vacua for a full nonperturbative
string theory, it would seem to contradict the expectation of
 \cite{Acharya:2006zw} that there are only a finite number of vacua
 compatible with any finite region of physical parameter space.
It would be interesting to study these solutions further
to ascertain whether they have a specific physical nonperturbative
instability.  Another interesting feature of the distribution of
solutions shown in Figure~\ref{f:gl1} is that it does not contain any
solutions with small $g, \Lambda$.  This stands in contrast to the
cases where both $P_2(\tau)$ and $P_3(\tau)$ have three real roots, as
we now discuss.

Although the above gauge-fixing procedure breaks down in the case
where all roots of $P_2 (\tau)$ and $P_3 (\tau)$ are real, we have
generated a representative sample of fluxes in this category up to
$N=20$ and numerically solved the equations of motion. It is then
possible to remove the modular redundancy by choosing the solution for
$\tau $ to lie within $\cal F $, and then imposing the other
gauge-fixing conditions. The resulting distribution of solutions is
shown in Figure~\ref{f:gl2}.  \FIGURE{
\epsfig{file=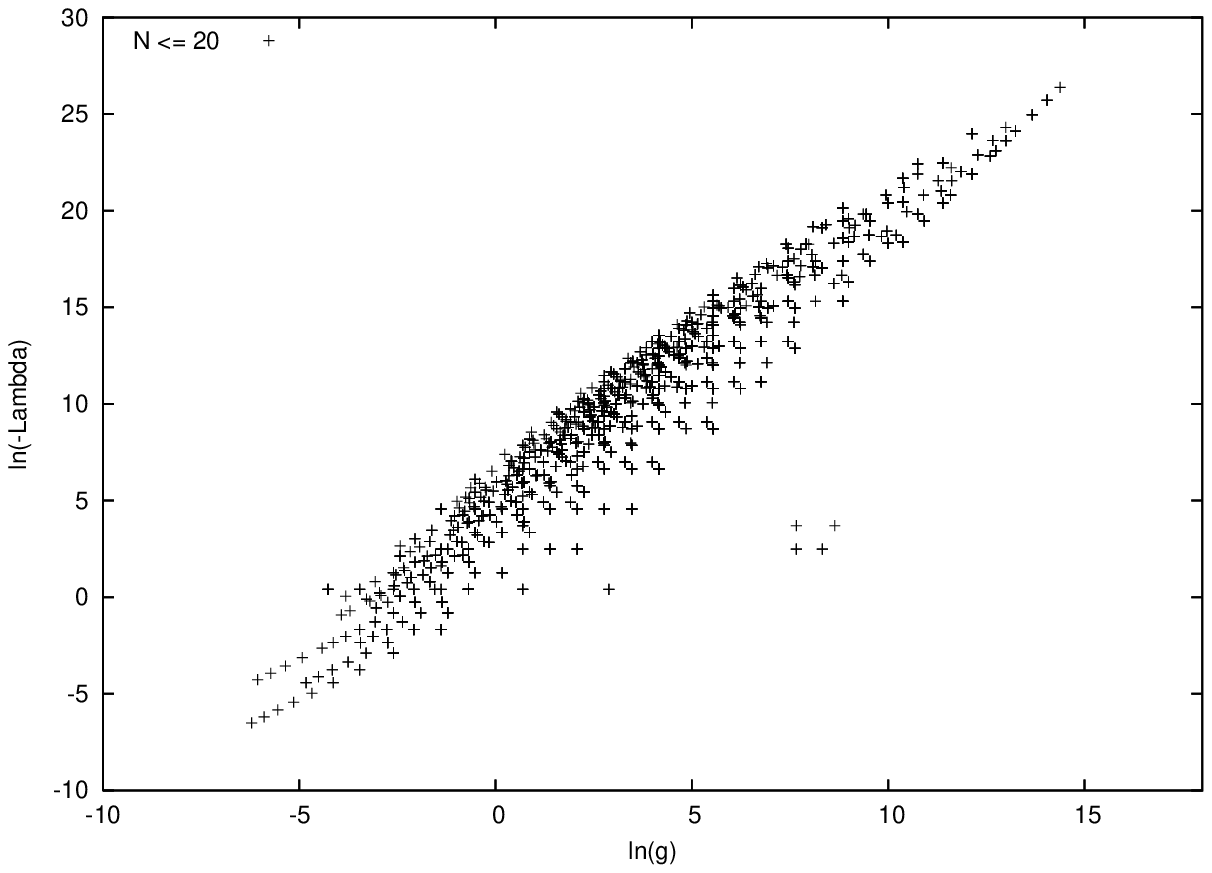,width=15cm}
\caption{\footnotesize Log-log plot of $g, \Lambda$ for 
  fluxes where $P_2, P_3$ have real roots up to $N =  20$}
\label{f:gl2}
} This distribution has the same broad features as that shown in
Figure~\ref{f:gl1}. In particular, we still see evidence for an
infinite number of distinct solutions in finite regions of $g$-$\Lambda$-space. The only major difference between this case and the
complex root case in Figure~\ref{f:gl1} is that here, much smaller
values of $g$ and $\Lambda$ appear.  Indeed, many of the points in
Figure~\ref{f:gl2} are members of the controllable families discussed in the
previous section.  It would be nice to have a clearer quantitative
understanding of why small values of $g$ and $ \Lambda$ seem so much
more difficult to realize in the case where $P_2 (\tau),\, P_3 (\tau)$
have complex roots.

\section{Conclusions}

The main goal of this paper has been to investigate supersymmetric
vacua of the ${\cal N}=1$ effective theory developed in \cite{stw}. We
find that generic supersymmetric solutions have the following
properties: all moduli are stabilized at tree level; the solutions are
not gauge-equivalent to geometric compactifications in any duality
frame; the string coupling and four-dimensional cosmological constant
are not particularly small; and there appear to be an infinite number of solutions in
finite regions of parameter space without vanishing or infinite
limits.  In addition, we have found infinite families of
vacua which have parametrically controllable string coupling and mass
scales.

The apparently infinite number of solutions in bounded regions of
parameter space is perhaps a bit surprising, and seems to conflict
with the intuition underlying recent speculations regarding finiteness
of the landscape of string vacua \cite{Acharya:2006zw, Ooguri-Vafa}.
Indeed, if there are an infinite number of valid string vacua in a
finite range of parameter space in effective field theory, with
sufficiently small (or negative) cosmological constants, it is difficult
to avoid an infinite transition rate into those vacua via vacuum decay
processes (though not impossible, if for example increasing a parameter such as a flux sufficiently quickly increases the height or width of the potential barrier separating these vacua from the physical vacuum).  It is therefore worthwhile to discuss a number of possible
reasons why the solutions we have found here may not represent true
supersymmetric string vacua.

To begin with, as we have discussed, it is quite possible that the
four-dimensional effective field theories studied here simply cannot
be lifted to a full string background, even with the inclusion of
other modes with masses comparable to those of the moduli.  Even if
these field theories can be lifted to geometric or nongeometric
backgrounds, given the difficulty of achieving modular invariance in
nongeometric asymmetric orbifold constructions \cite{ao2,Aoki:2004sm}
it is conceivable that the class of generalized flux compactifications
we consider here must obey unexpected constraints at the one-loop
level which would eliminate many or all of the tree-level solutions
given here.  It is also possible that we have missed a tree-level
constraint on the allowed fluxes, although it is somewhat difficult to
see how further constraints might arise.  Actually constructing string
backgrounds which incorporate nongeometric fluxes is probably the most
promising way of analyzing this set of possibilities 
further, and we leave this as an open problem.

Another possibility is that we have misidentified the modular
redundancy of the theory, and that many of the distinct solutions
which we find should in fact be considered to be equivalent
descriptions of the same solution.  We can
conceptually  separate two different ways in which the number of vacua becomes
large: first, the fractional axionic shifts which give potentially
large numbers of degenerate vacua with identical $g$ and $\Lambda$,
and second, the increasing density of distinct values of $g$ and
$\Lambda$ within a finite region as the maximum flux $N $ is
increased. There are no known symmetries which would identify these vacua, and our numerical work shows no evidence of
extra redundancies associated with unknown modular equivalences which
would reduce the number of inequivalent vacua to a finite number. It is conceivable, however, that such a symmetry exists, which would reduce the number of physically inequivalent solutions.

Yet another possibility is the destabilization of the
tree-level solutions by the inclusion of either $\alpha ^\prime $
effects or neglected momentum or winding modes.  If our solutions do
lift to full string backgrounds, we expect in general that $\alpha' $
effects will be significant.  In conventional flux compactifications,
one can often work in a large volume limit, where $\alpha^\prime$
corrections are highly suppressed; we are unable to do that here.
Indeed, in simple examples nongeometric fluxes typically stabilize
length scales of the compactification manifold at the string scale by
relating radii to their duals under monodromy, so we expect
$\alpha^\prime$ corrections to generically be important.  Finally, it
is possible that we should not allow the NS-NS fluxes to become
arbitrarily large, as such a flux configuration might lead to large
local backreaction, as we have discussed briefly at the end of Section
4.

Despite all these possible problems in extending the solutions we have
found to full-fledged string theory solutions, the generalized fluxes
appearing in the superpotential seem to capture essentially
topological features of the string compactification, and it seems
to us likely that the vacua found here do capture the main features of a new class of string compactifications.  Even
if these four-dimensional effective theories do not include all the
physics of the full string theory, they may point the way to a new and
interesting class of constructions of string vacua.  Certainly, the
question of which four-dimensional effective supergravity theories can
be lifted to complete string theories is a very important one which
deserves more attention at this time.  Whether this question is best
resolved by a top-down (landscape \cite{Susskind:2003kw}) approach of
attempting to characterize the most general structure of string theory
compactifications, or by a bottom-up (swampland
\cite{Ooguri:2006in,swampland}) approach of understanding which
four-dimensional theories have a good UV completion remains to be
seen.  In any case, a full characterization of the space of string
vacua which might be compared productively with experiment requires a
resolution of this question.  It is our hope that the solutions and
caveats which we have developed here may provide a useful set of test
cases to better understand the boundaries of the landscape.

\begin{center}
\bf{Acknowledgements}
\end{center}
\medskip

We would like to thank Michael Douglas, Henriette Elvang, Mariana
Gra\~{n}a, Nick Halmagyi, Albion Lawrence, Jan Louis, Shamit Kachru,
David Marks, Ilarion Melnikov, Ruben Minasian, Greg Moore, Sav Sethi,
Alessandro Tomasiello, Cumrun Vafa, and Brook Williams for helpful
conversations.  This work was supported in part by the DOE under
contract \#DE-FC02-94ER40818.  The research of research of WT was also
supported in part by the Stanford Institute for Theoretical Physics
(SITP).  The research of BW is additionally supported by NSF grant
PHY-00-96515.  JS and WT would like to think the KITP for hospitality
during part of this work.  WT would also like to thank Harvard
University for hospitality during part of this work.

\vspace*{0.2in}

\appendix

\section{Modular transformations}
\label{a:modular}

Here we give the transformation properties of the fluxes under the
modular transformations described in Section \ref{sec:eom}.

First, consider the shift of the axiodilaton $S\rightarrow S + n$.
This shift of the modulus is accompanied by a shift of the flux
parameters,
\beq
\label{eq:modular-begin}
a_i\rightarrow  a_i+ n b_i,
\eeq
with $b_i$, $c_i$ invariant, leaving the superpotential invariant under 
the shift.

Second, consider the shift $U\rightarrow U+n$.  Under this transformation 
the fluxes transform as
\barray
\nonumber
a_0 &\rightarrow & a_0-3 nc_0\\
a_1 &\rightarrow & a_1+ n (2 c_1-\tilde c_1)\\
\nonumber
a_2 &\rightarrow & a_2 + n (2c_2-\tilde c_2)\\
\nonumber
a_3 &\rightarrow & a_3-3 nc_3,
\earray
with $b_i$, $c_i$ invariant.

Finally, consider the shift $\tau\rightarrow\tau + n$, which is a
geometric transformation in type IIB.  The transformation laws for the
$a_i$, $b_i$ are, as usual,
\barray
\nonumber
a_0 &\rightarrow & a_0+3 a_1 n+3 a_2 n ^ 2+a_3 n ^ 3\\
a_1 &\rightarrow & a_1+2 na_2+ n ^ 2 a_3\\
\nonumber
a_2 &\rightarrow & a_2+ n a_3\\
\nonumber
a_3 &\rightarrow & a_3,
\earray
while the transformations of the $c_i$ are slightly more involved:
\barray
\nonumber
c_0 &\rightarrow & c_0-2 nc_1+ n\tilde{c}_1-2 n ^ 2c_2+ n ^ 2\tilde{c}_2+ n ^ 3 c_3\\
\nonumber
c_1 &\rightarrow & c_1-n\tilde{c}_2+ n c_2-n ^ 2 c_3\\
\tilde{c}_1 &\rightarrow &\tilde{c}_1-2 n c_2+ n ^ 2 c_3\\
\nonumber
c_2 &\rightarrow & c_2-nc_3\\
\nonumber
\tilde{c}_2 &\rightarrow &\tilde{c}_2+nc_3\\
\nonumber
c_3 &\rightarrow & c_3.
\earray
The separate transformations of $c_1$ and $\tilde{c}_1$ can be deduced
most easily from IIB, where the shift $\tau\rightarrow\tau + n$ is a
large diffeomorphism and the $c_i$ are all given by $Q ^{ab}_c$ on
different cycles.  By taking $Q$ to transform as a mixed tensor under
diffeomorphisms one may obtain the transformation rules above for the
integrated fluxes.

There is also an inversion symmetry $\tau\rightarrow-1/\tau$, under which
\barray
\nonumber
a_0 &\rightarrow & a_3\\
a_1 &\rightarrow &-a_2\\
\nonumber
a_2 &\rightarrow & a_1\\
\nonumber
a_3 &\rightarrow &-a_0
\earray
and similarly for the $b_i$; meanwhile,
\barray
\nonumber
c_0 &\rightarrow & c_3\\
\label{eq:modular-end}
c_1 &\rightarrow &-c_2\\
\nonumber
c_2 &\rightarrow & c_1\\
\nonumber
c_3 &\rightarrow &-c_0.
\earray
Under this transformation $W\rightarrow-(1/\tau ^ 3) W$, which
combined with the change in the K\"{a}hler potential \er{Kaehler}
leaves the scalar potential invariant.

The last factor in the modular group, as described in
Section~\ref{sec:eom}, is a $\bZ_2$ transformation which flips the
signs of all fluxes, leaving the moduli invariant.

One may check that all the constraints are preserved under the above modular transformations.

In Section 5 we fix in addition the global transformation which takes
$\tau, S, U\rightarrow-\bar\tau,-\bar S,-\bar U$.  Under this
transformation the fluxes which multiply odd powers of the moduli in
the superpotential switch signs,
\beq
\label{eq:global-flip}
a_1, a_3, b_0, b_2, c_0, c_2,\tilde c_2\rightarrow -a_1, -a_3, -b_0, -b_2, -c_0, -c_2,-\tilde c_2
\eeq
so that $W\rightarrow\bar W$.



\begin{thebibliography}{10}

\bibitem{flux-c} 
A.\ Strominger, ``Superstrings with torsion,''
  Nucl.\ Phys.\ B {\bf 274}, 253 (1986);
  J.~Polchinski and A.~Strominger,
  ``New Vacua for Type II String Theory,''
  Phys.\ Lett.\ B {\bf 388}, 736 (1996)
  {\tt hep-th/9510227};
  K.~Becker and M.~Becker,
  ``M-Theory on Eight-Manifolds,''
  Nucl.\ Phys.\ B {\bf 477}, 155 (1996)
  {\tt hep-th/9605053};
  J.~Michelson,
  ``Compactifications of type IIB strings to four dimensions with  non-trivial
  classical potential,''
  Nucl.\ Phys.\ B {\bf 495}, 127 (1997)
  {\tt hep-th/9610151}.



\bibitem{gvw}
  S.~Gukov, C.~Vafa and E.~Witten,
  ``CFT's from Calabi-Yau four-folds,''
  Nucl.\ Phys.\ B {\bf 584}, 69 (2000)
  [Erratum-ibid.\ B {\bf 608}, 477 (2001)]
  {\tt hep-th/9906070};
  T.~R.~Taylor and C.~Vafa,
  ``RR flux on Calabi-Yau and partial supersymmetry breaking,''
  Phys.\ Lett.\ B {\bf 474}, 130 (2000)
  {\tt hep-th/9912152}.
  
  \bibitem{drsgkp}
  K.~Dasgupta, G.~Rajesh and S.~Sethi,
  ``M theory, orientifolds and G-flux,''
  JHEP {\bf 9908}, 023 (1999)
  {\tt hep-th/9908088};
  S.~B.~Giddings, S.~Kachru and J.~Polchinski,
  ``Hierarchies from fluxes in string compactifications,''
  Phys.\ Rev.\ D {\bf 66}, 106006 (2002)
  {\tt hep-th/0105097}.
  
  \bibitem{glmw}
  S.~Gurrieri, J.~Louis, A.~Micu and D.~Waldram,
  ``Mirror symmetry in generalized Calabi-Yau compactifications,''
  Nucl.\ Phys.\ B {\bf 654}, 61 (2003)
  {\tt hep-th/0211102};
  S.~Gurrieri and A.~Micu,
  ``Type IIB theory on half-flat manifolds,''
  Class.\ Quant.\ Grav.\  {\bf 20}, 2181 (2003)
  {\tt hep-th/0212278}.

\bibitem{Hitchin}
N.\ Hitchin,
``Generalized Calabi-Yau manifolds,'' {\tt  math.dg/0209099}.

\bibitem{ss}
  J.~Scherk and J.~H.~Schwarz,
  ``How To Get Masses From Extra Dimensions,''
  Nucl.\ Phys.\ B {\bf 153}, 61 (1979).
  
  \bibitem{km}
  N.~Kaloper and R.~C.~Myers,
  ``The O(dd) story of massive supergravity,''
  JHEP {\bf 9905}, 010 (1999)
  {\tt hep-th/9901045}.
  
  \bibitem{x1}
  G.~Dall'Agata and S.~Ferrara,
  ``Gauged supergravity algebras from twisted tori compactifications with
  fluxes,''
  Nucl.\ Phys.\ B {\bf 717}, 223 (2005)
  {\tt hep-th/0502066}.
  
    \bibitem{stw}
  J.~Shelton, W.~Taylor and B.~Wecht,
  ``Nongeometric flux compactifications,''
  JHEP {\bf 0510}, 085 (2005)
  {\tt hep-th/0508133}.
  
  \bibitem{Dabholkar:2005ve}
  A.~Dabholkar and C.~Hull,
  ``Generalised T-duality and non-geometric backgrounds,''
  {\tt hep-th/0512005}.
  
  \bibitem{vp}
  P.~Fre, M.~Trigiante and A.~Van Proeyen,
  ``Stable de Sitter vacua from N = 2 supergravity,''
  Class.\ Quant.\ Grav.\  {\bf 19}, 4167 (2002)
  {\tt hep-th/0205119}.
  
  \bibitem{ferrara}
  G.~Dall'Agata and S.~Ferrara,
  ``Gauged supergravity algebras from twisted tori compactifications with
 fluxes,''
  Nucl.\ Phys.\ B {\bf 717}, 223 (2005)
  {\tt hep-th/0502066};
  G.~Dall'Agata, R.~D'Auria and S.~Ferrara,
  ``Compactifications on twisted tori with fluxes and free differential
  algebras,''
  Phys.\ Lett.\ B {\bf 619}, 149 (2005)
  {\tt hep-th/0503122};
  R.~D'Auria, S.~Ferrara and M.~Trigiante,
  ``Supersymmetric completion of M-theory 4D-gauge algebra from twisted tori
  and fluxes,''
  JHEP {\bf 0601}, 081 (2006)
  {\tt hep-th/0511158}.


  
  \bibitem{Buscher}
  T.~H.~Buscher,
  ``A Symmetry Of The String Background Field Equations,''
  Phys.\ Lett.\ B {\bf 194}, 59 (1987).

  \bibitem{x2}
  C.~M.~Hull,
  ``A geometry for non-geometric string backgrounds,''
  {\tt hep-th/0406102}.
  
  \bibitem{hre} 
  C.~M.~Hull and R.~A.~Reid-Edwards,
  ``Flux compactifications of string theory on twisted tori,''
  {\tt hep-th/0503114}.
  

\bibitem{global}
  C.~M.~Hull,
  ``Global aspects of T-duality, gauged sigma models and T-folds,''
  {\tt hep-th/0604178}.
  
\bibitem{Lawrence:2006ma}
  A.~Lawrence, M.~B.~Schulz and B.~Wecht,
  ``D-branes in nongeometric backgrounds,''
  {\tt hep-th/0602025}.
  
  
  
\bibitem{quantum}
  E.~Hackett-Jones and G.~Moutsopoulos,
  ``Quantum mechanics of the doubled torus,''
  {\tt hep-th/0605114};
 C.~M.~Hull,
  ``Doubled geometry and T-folds,''
  {\tt hep-th/0605149}.
  
    \bibitem{ao}
  K.~S.~Narain, M.~H.~Sarmadi and C.~Vafa,
  ``Asymmetric Orbifolds,''
  Nucl.\ Phys.\ B {\bf 288}, 551 (1987);
  S.~Hellerman, J.~McGreevy and B.~Williams,
  ``Geometric constructions of nongeometric string theories,''
  JHEP {\bf 0401}, 024 (2004)
  {\tt hep-th/0208174};
  A.~Dabholkar and C.~Hull,
  ``Duality twists, orbifolds, and fluxes,''
  JHEP {\bf 0309}, 054 (2003)
  {\tt hep-th/0210209}.
  
  \bibitem{ao2}
  A.~Flournoy, B.~Wecht and B.~Williams,
  ``Constructing nongeometric vacua in string theory,''
  Nucl.\ Phys.\ B {\bf 706}, 127 (2005)
  {\tt hep-th/0404217};
  A.~Flournoy and B.~Williams,
  ``Nongeometry, duality twists, and the worldsheet,''
  JHEP {\bf 0601}, 166 (2006)
  {\tt hep-th/0511126}.
  
\bibitem{Hellerman:2006tx}
  S.~Hellerman and J.~Walcher,
  ``Worldsheet CFTs for flat monodrofolds,''
  {\tt hep-th/0604191}.
  
    \bibitem{dgkt2}
  O.~DeWolfe, A.~Giryavets, S.~Kachru and W.~Taylor,
  ``Type IIA moduli stabilization,''
  {\tt hep-th/0505160}.
  
  \bibitem{Acharya:2006zw}
  B.~S.~Acharya and M.~R.~Douglas,
  ``A Finite Landscape?,''
  {\tt hep-th/0606212}.
  
  \bibitem{Ooguri:2006in}
  H.~Ooguri and C.~Vafa,
  ``On the geometry of the string landscape and the swampland,''
  {\tt hep-th/0605264}.

  \bibitem{Benmachiche:2006df}
    I.~Benmachiche and T.~W.~Grimm,
     ``Generalized N = 1 orientifold compactifications and the Hitchin
    functionals,''
    Nucl.\ Phys.\ B {\bf 748}, 200 (2006)
    {\tt hep-th/0602241}.
  
  \bibitem{bem}
  P.~Bouwknegt, J.~Evslin and V.~Mathai,
  ``On the topology and H-flux of T-dual manifolds,''
  Phys.\ Rev.\ Lett.\  {\bf 92}, 181601 (2004)
  {\tt hep-th/0312052}.
  

\bibitem{Frey:2002hf}
  A.~R.~Frey and J.~Polchinski,
  ``N = 3 warped compactifications,''
  Phys.\ Rev.\ D {\bf 65}, 126009 (2002)
  {\tt hep-th/0201029}.

  
  \bibitem{acfi}
  G.~Aldazabal, P.~G.~Camara, A.~Font and L.~E.~Ibanez,
  ``More dual fluxes and moduli fixing,''
  {\tt hep-th/0602089}; J.~Shelton and W.~Taylor, unpublished.
  
    \bibitem{dgkt1}
    O.~DeWolfe, A.~Giryavets, S.~Kachru and W.~Taylor,
  ``Enumerating flux vacua with enhanced symmetries,''
  JHEP {\bf 0502}, 037 (2005)
  {\tt hep-th/0411061}.

  
  
  \bibitem{kst}
  S.~Kachru, M.~B.~Schulz and S.~Trivedi,
  ``Moduli stabilization from fluxes in a simple IIB orientifold,''
  JHEP {\bf 0310}, 007 (2003)
  {\tt hep-th/0201028}.
 

  
  \bibitem{iib}
      S.~Kachru, M.~B.~Schulz and S.~Trivedi,
      ``Moduli stabilization from fluxes in a simple IIB orientifold,''
      JHEP {\bf 0310}, 007 (2003)
      {\tt hep-th/0201028};
  S.~Ferrara and M.~Porrati,
  ``N = 1 no-scale supergravity from IIB orientifolds,''
  Phys.\ Lett.\ B {\bf 545}, 411 (2002)
  {\tt hep-th/0207135};
    P.~K.~Tripathy and S.~P.~Trivedi,
  ``Compactification with flux on K3 and tori,''
  JHEP {\bf 0303}, 028 (2003)
  {\tt hep-th/0301139};
    R.~D'Auria, S.~Ferrara, F.~Gargiulo, M.~Trigiante and S.~Vaula,
  ``N = 4 supergravity Lagrangian for type IIB on T**6/Z(2) in presence of
 fluxes and D3-branes,''
  JHEP {\bf 0306}, 045 (2003)
  {\tt hep-th/0303049};
    A.~Font,
  ``$Z_N$ orientifolds with flux,''
  JHEP {\bf 0411}, 077 (2004)
  {\tt hep-th/0410206}.
  
    
  \bibitem{Kachru:2002sk}
  S.~Kachru, M.~B.~Schulz, P.~K.~Tripathy and S.~P.~Trivedi,
  ``New supersymmetric string compactifications,''
  JHEP {\bf 0303}, 061 (2003)
  {\tt hep-th/0211182}.
  
\bibitem{dkpz}
  J.~P.~Derendinger, C.~Kounnas, P.~M.~Petropoulos and F.~Zwirner,
  ``Superpotentials in IIA compactifications with general fluxes,''
  Nucl.\ Phys.\ B {\bf 715}, 211 (2005)
  {\tt hep-th/0411276};
  J.~P.~Derendinger, C.~Kounnas, P.~M.~Petropoulos and F.~Zwirner,
  ``Fluxes and gaugings: N = 1 effective superpotentials,''
  Fortsch.\ Phys.\  {\bf 53}, 926 (2005)
  {\tt hep-th/0503229}.


\bibitem{vz1}
  G.~Villadoro and F.~Zwirner,
  ``N = 1 effective potential from dual type-IIA D6/O6 orientifolds with
  general fluxes,''
  JHEP {\bf 0506}, 047 (2005)
  {\tt hep-th/0503169}.

\bibitem{cfi}
  P.~G.~Camara, A.~Font and L.~E.~Ibanez,
  ``Fluxes, moduli fixing and MSSM-like vacua in a simple IIA orientifold,''
  {\tt hep-th/0506066}.
  
  \bibitem{Gray:2006gn}
  J.~Gray, Y.~H.~He and A.~Lukas,
  ``Algorithmic algebraic geometry and flux vacua,''
  {\tt hep-th/0606122}.
  
\bibitem{stats}
  M.~R.~Douglas,
  ``The statistics of string / M theory vacua,''
  JHEP {\bf 0305}, 046 (2003)
  {\tt hep-th/0303194};
  F.~Denef and M.~R.~Douglas,
  ``Distributions of flux vacua,''
  JHEP {\bf 0405}, 072 (2004)
  {\tt hep-th/0404116};
  M.~R.~Douglas,
  ``Basic results in vacuum statistics,''
  Comptes Rendus Physique {\bf 5}, 965 (2004)
  {\tt hep-th/0409207}.
  


  \bibitem{ad}
    S.~Ashok and M.~R.~Douglas,
  ``Counting flux vacua,''
  JHEP {\bf 0401}, 060 (2004)
  {\tt hep-th/0307049}.

\bibitem{Ooguri-Vafa}
  H.~Ooguri and C.~Vafa,
  ``On the geometry of the string landscape and the swampland,''
  {\tt hep-th/0605264}.



  \bibitem{Aoki:2004sm}
  K.~Aoki, E.~D'Hoker and D.~H.~Phong,
  ``On the construction of asymmetric orbifold models,''
  Nucl.\ Phys.\ B {\bf 695}, 132 (2004)
  {\tt hep-th/0402134}.
  
  \bibitem{Susskind:2003kw}
  L.~Susskind,
  ``The anthropic landscape of string theory,''
  {\tt hep-th/0302219}.
  
  \bibitem{swampland}
    C.~Vafa,
  ``The string landscape and the swampland,''
  {\tt hep-th/0509212};
    N.~Arkani-Hamed, L.~Motl, A.~Nicolis and C.~Vafa,
  ``The string landscape, black holes and gravity as the weakest force,''
  {\tt hep-th/0601001};
  A.~Adams, N.~Arkani-Hamed, S.~Dubovsky, A.~Nicolis and R.~Rattazzi,
  ``Causality, analyticity and an IR obstruction to UV completion,''
  {\tt hep-th/0602178}.



\end{thebibliography}
\end{document}